\documentclass[aps,prd,eqsecnum,preprint,tightenlines,nofootinbib,showpacs]{revtex4-1}
\usepackage{graphicx,amsmath,latexsym}

\newcommand{\bea}{\begin{eqnarray}}
\newcommand{\eea}{\end{eqnarray}}
\newcommand{\be}{\begin{equation}}
\newcommand{\ee}{\end{equation}}
\newcommand{\ub}[1]{\underline{#1}}
\newcommand{\ob}[1]{\overline{#1}}
\newcommand{\Pminus}{{\cal P}^-}
\newcommand{\veck}{\vec{k}_\perp}
\newcommand{\veckp}{\vec{k}_\perp^{\,\prime}}

\begin{document}

\title{Application of a light-front coupled cluster method\footnote{Presented
by S.S. Chabysheva at LIGHTCONE 2011, 23-27 May 2011, Dallas.  To appear in
the proceedings.}}

\author{Sophia S. Chabysheva}
\author{John R. Hiller}
\affiliation{Department of Physics \\
University of Minnesota-Duluth \\
Duluth, Minnesota 55812}

\date{\today}

\begin{abstract}
As a test of the new light-front coupled-cluster method
in a gauge theory, we apply it to the nonperturbative
construction of the dressed-electron state in QED, for
an arbitrary covariant gauge, and compute the electron's
anomalous magnetic moment.  The construction
illustrates the spectator and Fock-sector independence
of vertex and self-energy contributions and indicates
resolution of the difficulties with uncanceled divergences
that plague methods based on Fock-space truncation.
\end{abstract}

\maketitle

\section{Introduction}
\label{sec:intro}

We wish to construct a nonperturbative solution for the
dressed-electron state in QED, as an eigenstate of
the light-front QED Hamiltonian, and use it to compute the
electron's anomalous magnetic moment.  Traditionally,
such a calculation is done as a truncated expansion of the
eigenstate in a Fock basis.  As discussed elsewhere
in these proceedings~\cite{hc}, truncation causes a number
of difficulties but can be avoided with use of the new
light-front coupled-cluster (LFCC) method~\cite{LFCClett}.
Here we give details of an application of the method begun
in \cite{hc}.  Our notation for light-cone 
coordinates~\cite{Dirac,DLCQreview} can also be found there.

The eigenvalue problem to solve  is
$\Pminus|\psi\rangle=\frac{M^2+P_\perp^2}{P^+}|\psi\rangle$.
The starting point of the LFCC method is to build the eigenstate
as $|\psi\rangle=\sqrt{Z}e^T|\phi\rangle$ from a valence state $|\phi\rangle$
and an operator $T$ that increases particle number but conserves
all relevant quantum numbers.  This leads to an
effective Hamiltonian $\ob{\Pminus}=e^{-T}\Pminus e^T$, which in
practice is computed with use of its Baker--Hausdorff expansion,
$\ob{\Pminus}=\Pminus+[\Pminus,T]+\frac12 [[\Pminus,T],T]+\cdots$.
The new eigenvalue problem
$\ob{\Pminus}|\phi\rangle=\frac{M^2+P_\perp^2}{P^+}|\phi\rangle$
is projected onto the valence and orthogonal sectors
\be
P_v\ob{\Pminus}|\phi\rangle=\frac{M^2+P_\perp^2}{P^+}|\phi\rangle, \;\;
(1-P_v)\ob{\Pminus}|\phi\rangle=0.
\ee

To have a finite system of equations, the operator $T$ and the
projection $1-P_v$ are truncated, with the latter chosen to
provide as many equations as are needed to determine the
functions in the truncated $T$.  This automatically truncates
the Baker--Hausdorff expansion of the effective Hamiltonian.
What is not truncated is the exponentiation of $T$ that builds
the eigenstate.  Thus the eigenstate retains all Fock states
consistent with the quantum numbers of the valence state.

The calculation of matrix elements requires some care to
avoid infinite sums over the Fock basis.
Consider the matrix element $\langle\psi_2|\hat O|\psi_1\rangle$
of an operator $\hat O$.  Define $|\psi_i\rangle=\sqrt{Z_i}e^{T_i}|\phi_i\rangle$, 
with $Z_i=1/\langle\phi_i|e^{T_i^\dagger}e^{T_i}|\phi_i\rangle$,
\be
\ob{O_i}=e^{-T_i}\hat O e^{T_i}
               =\hat{O}_i+[\hat{O}_i,T]+\frac12[[\hat{O}_i,T],T]+\cdots
\ee
and, to avoid the infinite sum in the denominator,
\be
\langle\widetilde\psi_i|=\langle\phi|\frac{e^{T_i^\dagger}e^T_i}
      {\langle\phi|e^{T_i^\dagger} e^{T_i}|\phi\rangle}
      =Z_i\langle\phi|e^{T_i^\dagger}e^{T_i}=\sqrt{Z_i}\langle\psi_i|e^{T_i} .
\ee
We then have
\be
\langle\psi_2|\hat O|\psi_1\rangle
    =\sqrt{Z_1/Z_2}\langle\widetilde\psi_2|\ob{O_2}e^{-T_2}e^{T_1}|\phi_1\rangle
=\sqrt{Z_2/Z_1}\langle\widetilde\psi_1|\ob{O_1^\dagger}e^{-T_1}e^{T_2}|\phi_2\rangle^* .
\ee
Therefore, we can compute the matrix element as
\be
\langle\psi_2|\hat O|\psi_1\rangle
  =\sqrt{\langle\tilde\psi_2|\ob{O_2}e^{-T_2}e^{T_1}|\phi_1\rangle
     \langle\widetilde\psi_1|\ob{O_1^\dagger}e^{-T_1}e^{T_2}|\phi_2\rangle^*} .
\ee
In the diagonal case, this reduces to the form discussed in \cite{hc}:
\be
\langle\psi|\hat O|\psi\rangle=\langle\widetilde\psi|\ob{O}|\phi\rangle.
\ee

\section{Application to QED}

As a test of the LFCC method, we consider the dressed-electron state
truncated to exclude positrons but retaining an infinite number of
photons.  From this state, we compute the anomalous moment from a
spin-flip matrix element of the current.  The different spin states,
for both the physical and PV eigenstates, are computed simultaneously
with a single $T$ operator; the matrix element can then be computed
from the simpler diagonal form.

\subsection{Effective Hamiltonian}

We use a QED Lagrangian regulated by Pauli--Villars (PV) fields~\cite{ArbGauge}
\be
{\cal L} =  \sum_{i=0}^2 (-1)^i \left[-\frac14 F_i^{\mu \nu} F_{i,\mu \nu} 
         +\frac12 \mu_i^2 A_i^\mu A_{i\mu} 
         -\frac12 \zeta \left(\partial^\mu A_{i\mu}\right)^2\right]
+ \sum_{i=0}^2 (-1)^i \bar{\psi_i} (i \gamma^\mu \partial_\mu - m_i) \psi_i 
  - e \bar{\psi}\gamma^\mu \psi A_\mu,
\ee 
where the interaction is written in terms of null fields
\be \label{eq:NullFields}
  \psi =  \sum_{i=0}^2 \sqrt{\beta_i}\psi_i, \;\;
  A_\mu  = \sum_{i=0}^2 \sqrt{\xi_i}A_{i\mu}, \;\;
  F_{i\mu \nu} = \partial_\mu A_{i\nu}-\partial_\nu A_{i\mu}
\ee
and the coupling coefficients are constrained by
\be
\xi_0=1, \;\;
\sum_{i=0}^2(-1)^i\xi_i=0, \;\;
\beta_0=1, \;\;
\sum_{i=0}^2(-1)^i\beta_i=0,
\ee
with a zero index for physical fields and nonzero for PV fields.
Further, we require chiral symmetry restoration~\cite{ChiralLimit}
and zero photon mass~\cite{VacPol}, to fix $\xi_2$ and $\beta_2$.
The theory is quantized with a light-front analog
of the Stueckelberg method~\cite{Stueckelberg}.  The light-front
Hamiltonian, without positron contributions, is~\cite{hc,ArbGauge}
\bea
\Pminus &=&   \sum_{i,s}\int d\ub{p}
      \frac{m_i^2+p_\perp^2}{p^+}(-1)^i
          b_{i,s}^\dagger(\ub{p}) b_{i,s}(\ub{p})
           +\sum_{l,\lambda}\int d\ub{k}
          \frac{\mu_{l\lambda}^2+k_\perp^2}{k^+}(-1)^l\epsilon^\lambda
             a_{l\lambda}^\dagger(\ub{k}) a_{l\lambda}(\ub{k}) \\
&& +\sum_{ijls\sigma\lambda}\int dy d\vec{k}_\perp 
   \int\frac{d\ub{p}}{\sqrt{16\pi^3p^+}}
             \left\{h_{ijl}^{\sigma s\lambda}(y,\vec{k}_\perp)\right.  \nonumber \\
&& \left. \times  a_{l\lambda}^\dagger(yp^+,y\vec{p}_\perp+\vec{k}_\perp)
   b_{js}^\dagger((1-y)p^+,(1-y)\vec{p}_\perp-\vec{k}_\perp)b_{i\sigma}(\ub{p})  + \mbox{H.c.}\right\},  \nonumber
\eea
with the $h_{ijl}^{\sigma s\lambda}$ being known functions and 
$\epsilon=(-1,1,1,1)$ the metric signature for the physical photon.

\begin{figure}
\centering
  \includegraphics[width=12cm]{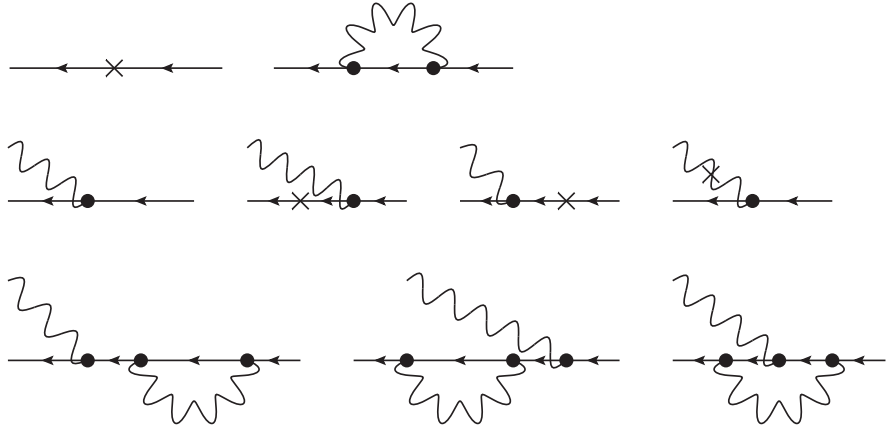}
\caption{Graphical representation of the terms of the effective Hamiltonian
in Eq.~(\ref{eq:EffH}) of the text. Each graph represents an operator that
annihilates an electron and creates either a single electron or an electron
and photon.  The crosses indicate light-front kinetic-energy contributions.}
\label{fig:fig1}
\end{figure}

We truncate the $T$ operator to just simple photon emission from
a fermion:
\be
T=\sum_{ijls\sigma\lambda}\int dy d\vec{k}_\perp 
   \int\frac{d\ub{p}}{\sqrt{16\pi^3}}\sqrt{p^+} t_{ijl}^{\sigma s\lambda}(y,\vec{k}_\perp)
a_{l\lambda}^\dagger(yp^+,y\vec{p}_\perp+\vec{k}_\perp)
   b_{js}^\dagger((1-y)p^+,(1-y)\vec{p}_\perp-\vec{k}_\perp)b_{i\sigma}(\ub{p}) .
\ee
The effective Hamiltonian, excluding terms that annihilate the valence
state of a single electron, becomes
\bea  \label{eq:EffH}
\ob{\Pminus} &=& \sum_{ijs}\int d\ub{p}(-1)^i
      \left[\delta_{ij}\frac{m_i^2+p_\perp^2}{p^+}+\frac{I_{ji}}{p^+}\right]
          b_{j,s}^\dagger(\ub{p}) b_{i,s}(\ub{p})   \\
&& +\sum_{ijls\sigma\lambda}\int dy d\vec{k}_\perp 
   \int\frac{d\ub{p}}{\sqrt{16\pi^3p^+}}
          \left\{h_{ijl}^{\sigma s\lambda}(y,\vec{k}_\perp)
            +\frac12 V_{ijl}^{\sigma s\lambda}(y,\vec k_\perp)\right.  \nonumber \\
&& +\left[\frac{m_j^2+k_\perp^2}{1-y}
                                +\frac{\mu_{l\lambda}^2+k_\perp^2}{y}-m_i^2\right]
                                t_{ijl}^{\sigma s\lambda}(y,\vec k_\perp) \nonumber \\
&& \left.
    +\frac12\sum_{i'}\frac{I_{ji'}}{1-y}t_{ii'l}^{\sigma s\lambda}(y,\vec k_\perp)
              -\sum_{j'}(-1)^{i+j'}t_{j'jl}^{\sigma s\lambda}(y,\vec k_\perp)I_{j'i}\right\}
                             \nonumber \\
&& \times  a_{l\lambda}^\dagger(yp^+,y\vec{p}_\perp+\vec{k}_\perp)
   b_{js}^\dagger((1-y)p^+,(1-y)\vec{p}_\perp-\vec{k}_\perp)b_{i\sigma}(\ub{p}), \nonumber
\eea
with the self-energy contribution 
\be
I_{ji}=(-1)^i\sum_{i'ls\lambda}(-1)^{i'+l}\epsilon^\lambda
        \int \frac{dy dk_\perp^2}{16\pi^3 p^+}
        h_{ji'l}^{\sigma s\lambda*}(y,\vec k_\perp) 
        t_{ii'l}^{\sigma s\lambda}(y,\vec k_\perp),
\ee
and the vertex loop correction
\bea
V_{ijl}^{\sigma s\lambda}(y,\veck)
&=&\sum_{i'j'l's'\sigma'\lambda'}(-1)^{i'+l'+j'}\epsilon^{\lambda'}
  \int \frac{dy' d\veckp}{16\pi^3}
  \frac{\theta(1-y-y')}{(1-y')^{1/2}(1-y)^{3/2}}  \\
  && \times
  h_{jj'l'}^{ss'\lambda'*}(\frac{y'}{1-y},\veckp+\frac{y'}{1-y}\veck)
  t_{i'j'l}^{\sigma' s'\lambda}(\frac{y}{1-y'},\veck+\frac{y}{1-y'}\veckp)
  t_{ii'l'}^{\sigma\sigma'\lambda'}(y',\veckp).  \nonumber
\eea
A graphical representation of the terms in $\ob{\Pminus}$ is
given in Fig.~\ref{fig:fig1}.

\subsection{Eigenvalue problems}

Because $\ob{\Pminus}$ is not Hermitian, we have both right and
left-hand valence states
\be
|\phi^\pm(\ub{P})\rangle=\sum_i z_i b_{i\pm}^\dagger(\ub{P})|0\rangle, \;\;
\langle\widetilde\phi^\pm(\ub{P})|=\langle0|\sum_i \tilde{z}_i b_{i\pm}(\ub{P}) .
\ee
The projections of the right and left-hand eigenvalue problems
onto the valence sector,
\be
P_v \ob{\Pminus}P_v|\phi^\pm(\ub{P})\rangle=\frac{M^2+P_\perp^2}{P^+}|\phi^\pm(\ub{P})\rangle
\ee
and
\be 
(P_v \ob{\Pminus}P_v)^\dagger|\widetilde\phi^\pm(\ub{P})\rangle
    =\frac{M^2+P_\perp^2}{P^+}|\widetilde\phi^\pm(\ub{P})\rangle ,
\ee
yield
\be
m_i^2 z_{ai}^\pm +\sum_j I_{ij} z_{aj}^\pm = M_a^2 z_{ai}^\pm
\ee
and
\be
m_i^2 \tilde{z}_{ai}^\pm +\sum_j (-1)^{i+j}I_{ji} \tilde{z}_{aj}^\pm 
     = M_a^2 \tilde{z}_{ai}^\pm,
\ee
with $a=0,1$ and $M_a$ the $a$th eigenmass.
The valence eigenvectors are orthonormal and complete,
in the following sense:
\be
\sum_i (-1)^i \tilde{z}_{ai}^\pm z_{bi}^\pm=(-1)^a \delta_{ab}, \;\;
\sum_a (-1)^a z_{ia}^\pm \tilde{z}_{ja}^\pm = (-1)^i \delta_{ij} .
\ee

Projection of the right-hand eigenvalue problem onto $|e\gamma\rangle$,
orthogonal to $|\phi\rangle$, gives
\bea
&&\sum_i(-1)^i z_{ai}^\pm\left\{h_{ijl}^{\pm s\lambda}(y,\veck)
+\frac12 V_{ijl}^{\pm s\lambda}(y,\veck)
+\left[\frac{m_j^2+k_\perp^2}{1-y}+\frac{\mu_{l\lambda}^2+k_\perp^2}{y}-m_i^2\right]
                 t_{ijl}^{\pm s\lambda}(y,\veck) \right. \\
&& \left. +\frac12\sum_{i'} \frac{I_{ji'}}{1-y} t_{ii'l}^{\pm s\lambda}(y,\veck)
-\sum_{j'}(-1)^{i+j'}t_{j'jl}^{\pm s\lambda}(y,\veck)I_{j'i} \right\}=0 .  \nonumber
\eea
To partially diagonalize in flavor, we define
\be
C_{abl}^{\pm s\lambda}(y,\veck)
  =\sum_{ij}(-1)^{i+j}z_{ai}^\pm \tilde{z}_{bj}^\pm t_{ijl}^{\pm s\lambda}(y,\veck) .
\ee
With analogous definitions for $H$, $I$, and $V$, we have
\be
\left[M_a^2-\frac{M_b^2+k_\perp^2}{1-y}-\frac{\mu_{l\lambda}^2+k_\perp^2}{y}\right]
   C_{abl}^{\pm s\lambda}(y,\veck) 
=H_{abl}^{\pm s\lambda}(y,\veck) 
 +\frac12\left[V_{abl}^{\pm s\lambda}(y,\veck)
      -\sum_{b'}\frac{I_{bb'}}{1-y}C_{ab'l}^{\pm s\lambda}(y,\veck)\right],
\ee
to be solved simultaneously with the valence-sector equations, which depend
on $C$ (or $t$) through the self-energy matrix $I$.
Notice that the physical mass $M_b$ has replaced the bare mass in
the kinetic energy term, without use of sector-dependent 
renormalization~\cite{Wilson,hb,Karmanov,SecDep}.

The dual to the left-hand eigenstate
$\langle\widetilde\psi^\pm(\ub{P})|=\sqrt{Z}\langle\psi^\pm(\ub{P})|e^T$
is a right eigenstate of $\ob{\Pminus}^\dagger$:
\be
\ob{\Pminus}^\dagger|\tilde\psi^\pm(\ub{P})\rangle
=e^{T^\dagger}\Pminus e^{-T^\dagger} \sqrt{Z}e^{T^\dagger}|\psi^\pm(\ub{P})\rangle
=\frac{M^2+P_\perp^2}{P^+}|\tilde\psi^\pm(\ub{P})\rangle .
\ee
It is normalized such that
\be
\langle\tilde\psi^\pm(\ub{P'})|\phi(\ub{P})\rangle
=\sqrt{Z}\langle\psi^\pm(\ub{P'})|e^T|\phi^\pm(\ub{P})\rangle
=\langle\psi^\pm(\ub{P'})|\psi(\ub{P})\rangle
=\delta(\ub{P'}-\ub{P}) .  \nonumber
\ee
For the dressed electron, we construct this state as
\be
|\tilde\psi_a^\pm(\ub{P})\rangle=|\tilde\phi_a^\pm(\ub{P})\rangle
    + \sum_{jls\lambda}\int dy d\veck \sqrt{\frac{P^+}{16\pi^3}}
       l_{ajl}^{\pm s\lambda}(y,\veck)
  a_{l\lambda}^\dagger(yP^+,y\vec{P}_\perp+\veck)
       b_{js}^\dagger((1-y)P^+,(1-y)\vec{P}_\perp-\veck)|0\rangle .
\ee 
We then flavor-diagonalize the left-hand wave functions 
\be
D_{abl}^{\pm s\lambda}(y,\veck)\equiv\sum_j(-1)^j z_{bj}^s l_{ajl}^{\pm s\lambda}(y,\veck),
\ee
to obtain
\be
\left[M_a^2-\frac{M_b^2+k_\perp^2}{1-y}
    -\frac{\mu_{l\lambda}^2+k_\perp^2}{y}\right]
        D_{abl}^{\sigma s\lambda}(y,\veck) 
 =\tilde{H}_{abl}^{\sigma s\lambda}(y,\veck) 
                  +W_{abl}^{\sigma s\lambda}(y,\veck)
 -\sum_{b'} J_{b'a}^\sigma \tilde{H}_{b'bl}^{\sigma s\lambda}(y,\veck),
\ee
where 
\be
\widetilde{H}_{abl}^{\sigma s\lambda}(y,\veck)
  =\sum_{ij}(-1)^{i+j}\tilde{z}_{ai} z_{bj} h_{ijl}^{\sigma s\lambda}(y,\veck) ,
\ee
$W_{abl}^{\sigma s\lambda}$ is a vertex-correction analog of
$V_{abl}^{\sigma s\lambda}$, though linear in $D$, and $J_{ba}^\sigma$
is a self-energy analog of $I_{ba}$.
Solutions for $M_a$, $z_{ai}^\sigma$, $\tilde{z}_{ai}^\sigma$, 
and $C_{abl}^{\sigma s \lambda}$ are used as input.

\subsection{Anomalous moment}

We compute the anomalous moment $a_e$ from the spin-flip
matrix element of the current 
$J^+=\overline{\psi}\gamma^+\psi$~\cite{BrodskyDrell},
coupled to a photon of momentum $q$
in the Drell--Yan ($q^+=0$) frame~\cite{DrellYan}:
\be
16\pi^3\langle\psi_a^\sigma(\ub{P}+\ub{q})|J^+(0)|\psi_a^\pm(\ub{P})\rangle
=2\delta_{\sigma\pm}F_1(q^2)\pm\frac{q^1\pm iq^2}{M_a}\delta_{\sigma\mp}F_2(q^2).
\ee
In limit of infinite PV masses, and with $M_0=m_e$ the electron mass, we have
\bea
F_1(q^2)&=&\frac{1}{\cal N}\left[1+\sum_s\int\frac{dy d\veck}{16\pi^3}
\left\{\sum_{\lambda=1}^2 
  l_{000}^{\pm s\lambda *}(y,\veck-y\vec{q}_\perp)t_{000}^{\pm s\lambda}(y,\veck)
                      \right.\right.\\
 && \left. \left. \rule{1.5in}{0mm}
 -\sum_{\lambda=0}^3\epsilon^\lambda
    l_{000}^{\pm s \lambda *}(y,\veck) t_{000}^{\pm s \lambda}(y,\veck)\right\}\right]
 \nonumber
\eea
and
\be
F_2(q^2)=\pm\frac{2m_e}{q^1\pm iq^2}
\sum_s\sum_{\lambda=1}^2\int\frac{dy d\veck}{16\pi^3} 
    l_{000}^{\mp s\lambda *}(y,\veck-y\vec{q}_\perp)
                         t_{000}^{\pm s\lambda}(y,\veck)/{\cal N},
\ee
with
\be
{\cal N}=1-\sum_s\sum_{\lambda=0,3}\epsilon^\lambda
   \int\frac{dy d\veck}{16\pi^3} l_{000}^{\pm s\lambda*}(y,\veck)
          t_{000}^{\pm s\lambda}(y,\veck).
\ee
A second term is absent in $F_2$ because $l$ and $t$ are orthogonal
for opposite spins.  

The $q^2\rightarrow0$ limit can be taken, to find $F_1(0)=1$ and
\be
a_e=F_2(0)=\pm \frac{m_e}{{\cal N}}\sum_s\sum_{\lambda=1,2}\epsilon^\lambda
\int \frac{dy d\veck}{16\pi^3}
y l_{000}^{\mp s\lambda *}(y,\veck)
\left(\frac{\partial}{\partial k^1}\mp i\frac{\partial}{\partial k^2}\right)
t_{000}^{\pm s\lambda}(y,\veck).
\ee
As a check, we can consider a perturbative solution 
\be
t_{000}^{\sigma s\lambda}
=l_{000}^{\sigma s\lambda}
=h_{000}^{\sigma s\lambda}
/\left[m_e^2-\frac{m_e^2+k_\perp^2}{1-y}
    -\frac{\mu_{l\lambda}^2+k_\perp^2}{y}\right].
\ee
Substitution into the expression for $a_e$ gives
immediately the Schwinger result~\cite{Schwinger} of $\alpha/2\pi$,
in the limit of zero photon mass, for any covariant gauge.
A complete calculation requires a numerical solution of
the eigenvalue problems for the left and right-hand wave 
functions.

\section{Summary} \label{sec:summary}

The LFCC method provides the means to solve for eigenstates
of light-front Hamiltonians without truncation of Fock space
and includes techniques for computing physical observables
from matrix elements.  The illustration given here shows
how the method is applied to a calculation of the dressed-electron
eigenstate of QED, including determination of the anomalous
moment.  A first-order perturbative solution yields the standard Schwinger
term.  A complete solution requires numerical techniques; work
on this is in progress.  The analysis and calculation is 
systematically improvable through the addition of terms
to the exponentiated operator $T$. In particular, for the
dressed electron, a term that adds electron-positron loops
to the calculation can be included.  A perturbative solution
would then match ordinary perturbation theory to order $\alpha^2$;
the nonperturbative solution provides a partial resummation
of contributions from higher orders.

\acknowledgments
This work was supported in part by the US Department of Energy
through Contract No.\ DE-FG02-98ER41087
and by the Minnesota Supercomputing Institute through
grants of computing time.

\end{document}